\documentstyle[preprint,revtex]{aps}
\renewcommand{\Re}{\mathop{\rm Re}}	 
\renewcommand{\Im}{\mathop{\rm Im}}	 
\begin{document}
\draft
%
%
\begin{title}
Tunneling and the Spectrum of the Potts Model
\end{title}
\author{Jing-Dong Wang and Carleton DeTar}
\begin{instit}
Department of Physics, University of Utah, Salt Lake City, UT 84112
\end{instit}
\receipt{}
\begin{abstract}
The three-dimensional, three-state Potts model is studied as a
paradigm for high temperature quantum chromodynamics.  In a high
statistics numerical simulation using a Swendson-Wang algorithm, we
study cubic lattices of dimension as large as $64^3$ and measure
correlation functions on long lattices of dimension $20^2\times 120$
and $30^2\times 120$.  These correlations are controlled by the
spectrum of the transfer matrix.  This spectrum is studied in the
vicinity of the phase transition.  The analysis classifies the
spectral levels according to an underlying $S_3$ symmetry.  Near the
phase transition the spectrum agrees nicely with a simple
four-component hamiltonian model.  In the context of this model, we
find that low temperature ordered-ordered interfaces nearly always
involve a disordered phase intermediate.  We present a new spectral
method for determining the surface tension between phases.
\end{abstract}
\section{Introduction}

The three-dimensional, three-state Potts model has long been studied
as a paradigm for the phase structure of quantum chromodynamics (QCD)
at high temperature in the heavy quark limit\cite{ref:pottsQCD}. It
has been found that at zero magnetic field, the Potts model has a weak
first order phase transition, separating a low-temperature phase that
breaks the Z(3) symmetry and a high-temperature phase in which the
symmetry is restored\cite{ref:order}. The underlying Z(3) symmetry
requires that at low temperature there be three ordered or broken
symmetry phases with the same free energy.  In a finite volume system
the behavior of the theory in the vicinity of the phase transition is
complicated by tunneling among four phases: the three ordered phases
and the disordered (symmetry restored) phase.

The Potts model with no magnetic field corresponds to QCD with
infinitely heavy quarks---in effect, without any dynamical quarks.  A
recent study of QCD without quarks by the APE group found that the
correlation length of the system appears to grow as the physical
volume is increased, suggesting an infinite correlation length in the
infinite volume limit, a characteristic of a continuous phase
transition\cite{ref:APE}. Doubts were raised that the phase transition
is first order. The APE study differed from other contemporary
work\cite{ref:Columbia} in measuring correlation lengths on lattices
with one long dimension.  Two more recent high statistics studies with
the Potts model and with QCD, using different indicators of the order
of the phase transition, have reconfirmed the first order character of
the phase transition in both models\cite{ref:Karsch,ref:Fukugita}.
These more recent studies used a finite-size scaling analysis of
susceptibilities to demonstrate a first order transition.  The now
widely suspected explanation for the confusion over correlation
lengths is that finite volume tunneling among the Z(3)-equivalent
ordered phases introduces a correlation related to the typical domain
size of those phases.  This correlation length does indeed become
infinite in an infinite volume system, as a natural consequence of the
first order character of the phase transition.

To test this suspected explanation and to develop new insights into
the effects of tunneling, we have carried out a new study of the
three-state, three-dimensional Potts model on lattices with one long
<dimension\cite{ref:wang}. Our study emphasizes the determination of
the spectrum of the transfer matrix.  We demonstrate explicitly how
tunneling modifies the spectrum and, as a consequence, the correlation
lengths in the vicinity of the phase transition.  Our study parallels
work done with the four-dimensional Ising model by Jansen et
al\cite{ref:Montvay}.  Our work makes it possible to understand the
results of the several
references\cite{ref:APE,ref:Columbia,ref:Karsch,ref:Fukugita} in a
common framework.  In the next section we discuss the $S_{3}$ symmetry
of the transfer matrix, develop a phenomenological four-component
model for tunneling, and introduce the formulas needed for obtaining
the spectrum.  In Sec.~\ref{sec:NumSim} we present results of the
simulation.  We show that the spectrum and interface statistics agree
well with predictions of the four-component model.  We obtain the
surface tension from the spectral splittings.  In the final section we
state our conclusions.

\section{Phenomenology of the Potts Model}

\subsection{Transfer Matrix and Symmetries}

The three-dimensional, three-state Potts model is a classical spin
system with one spin $s_i$ on each site $i = (x,y,z)$ of a cubic
lattice.  Let the lattice dimension be $L_x \times L_y \times L$.
Spins take on the values $s_i \in Z(3) = \{1, \exp(\pm2\pi i/3)\}$.
The partition function of the Potts model at zero magnetic field is
given by
\begin{equation}
  Z(\beta) = \sum_{s_i}\exp\left(-\beta \sum_{(i\mu)} \Re s^*_i
s_{i+\hat\mu}\right),
\end{equation}
where $i+\hat\mu$ is the nearest neighbor site in the positive
$\hat\mu$ direction and the sum $s_i$ is over all configurations of
spins.  Let the spins on a plane of constant lattice coordinate $z$ be
denoted $S_z = \{s_{x,y,z} | x \in 1,\ldots{},L_x, y \in
1,\ldots{},L_y\}$.  Then, as is well known, the partition function can
be written as the trace of the transfer matrix raised to the power
$L$:
\begin{equation}
   Z(\beta) = \mathop{\rm Tr} T^L, \label{eq:part_func}
\end{equation}
where
\begin{equation}
  \left\langle S_z\right| T \left| S_{z+1}\right\rangle = \exp\left(
-\beta \Re s^*_{(x,y,z)} s_{(x,y,z+1)} -\beta \sum_{\mu=1,2}\Re
s^*_{(x,y,z)} s_{(x,y,z)+\hat\mu} \right).
\end{equation}
The transfer matrix can be regarded as a quantum mechanical operator
acting on a Hilbert space of states described by configurations of
spins arranged on the $(x,y)$ lattice plane.  All operators of present
interest are local, and can be expressed as a function of the spins.
The expectation value of an operator $O$ on this ensemble is expressed
in terms of the transfer matrix as
\begin{equation}
  \left\langle O\right\rangle = \mathop{\rm Tr} (T^L O)/\mathop{\rm
Tr} T^L.
\end{equation}
Local operators can be regarded as depending on a specific lattice
plane $z$.  The correlation between local operators $A(z)$ and
$B(z^\prime)$ with $z^\prime > z$ is expressed as
\begin{equation}
  \left\langle A(z)B(z^\prime)\right\rangle = \mathop{\rm Tr} (T^z A
T^{z^\prime - z} B T^{L - z^\prime}) / \mathop{\rm Tr} T^L.
\end{equation}
A hamiltonian matrix $H$ is defined in terms of the transfer matrix so
that
\begin{equation}
  T = \exp(-H).  \label{eq:defT}
\end{equation}
Let us write the spectral decomposition of the transfer matrix in
terms of the eigenvalues $E_n$ of the hamiltonian $H$ as
\begin{equation}
  T = \sum_n \left| n\right\rangle \exp(-E_n)\left\langle n\right|.
\end{equation}
Let $n=0$ denote the ground state.  The correlation between two local
operators $A$ and $B$ can be written in terms of the spectrum as
follows:
\begin{equation}
    \left\langle A(z)B(z^\prime)\right\rangle =
\frac{\sum_{mn}\left\langle m\right|A\left| n\right\rangle
\left\langle n\right|B\left| m\right\rangle \exp[-(E_n-E_m)(z^\prime -
z)]\exp(-E_mL)} {\sum_m \exp(-E_m)} \label{eq:corr}
\end{equation}
In this way correlations between local operators give information
about the spectrum of the transfer matrix.  (Of course, only the
energy level differences $(E_n - E_0)$ have physical significance.)

The Potts model is symmetric under global transformations of the
three-fold permutation group $S_3$.  These transformations are
generated by
\begin{equation}
   s_i \rightarrow e^{\pm2\pi i/3} s_i \ \ ; \ \ s_i \rightarrow s^*_i
\ \ \ \forall_i.
\end{equation}
Therefore each of the eigenstates of $H$ and each of the operators of
interest $O$ can be classified according to the three irreducible
representations $S$ (symmetric), $A$ (antisymmetric) and $M$ (mixed,
two-dimensional).  For example, the spin operator, itself, belongs to
$M$, the operator $|s|^2$ belongs to $S$, and the operator $\Re s_i
\Im s_j - \Im s_i \Re s_j$ belongs to $A$.

\subsection{A Simple Model} \label{sec:model}

To begin with the classification of states, consider first the extreme
case $\beta \rightarrow \infty$.  As is well known, the statistical
ensemble reduces to three configurations of equal weight, with all
spins aligned in either of the three Z(3) directions.  In the
Hamiltonian language the ground state of $H$ at infinite $\beta$ is
three-fold degenerate, with all spins on the $(x,y)$ plane aligned in
one of the three directions.  Call these three states $\left|
1\right\rangle $, $\left| 2\right\rangle $, and $\left| 3\right\rangle
$.  These states are related by a Z(3) transformation $R$ as follows:
\begin{equation}
  \left| 2\right\rangle = R\left| 1\right\rangle \ \ \ \left|
3\right\rangle = R\left| 2\right\rangle .  \label{eq:Zrotation}
\end{equation}
At the other extreme $\beta = 0$ the statistical ensemble contains all
spin configurations with equal weight, and the ground state of the
hamiltonian is not degenerate.  Call it $\left| 0\right\rangle $.
These states have finite $\beta$ counterparts.

Beginning from these extremes, we introduce a phenomenological model
for the spectrum at intermediate $\beta$ in a finite volume system.
As $\beta$ is decreased from infinity, mixing between the three
degenerate states occurs.  Because of the $S_3$ symmetry, the
Hamiltonian matrix must be approximately of the form
\begin{equation}
\left(
  \begin{array}{ccc}
    0 \ \ & \epsilon \ \ & \epsilon \\ \epsilon \ \ & 0 \ \ & \epsilon
\\ \epsilon \ \ & \epsilon \ \ & 0
  \end{array}
\right).
\end{equation}
As a result of mixing, the degeneracy of the three states is lifted,
giving rise to a symmetric ground state $\left| 0s\right\rangle =
\left| 1\right\rangle + \left| 2\right\rangle + \left| 3\right\rangle
$ and a twofold degenerate mixed-symmetry state $\left|
0m\right\rangle $ of slightly higher energy.  The mixing parameter
$\epsilon$ depends on the transverse area $L_xL_y$.  Since mixing
between the degenerate vacuums requires a rearrangement of the spins
over the entire $(x,y)$ plane, it is plausible that the dependence is
\begin{equation}
 \epsilon = \exp(-\beta L_x L_y\alpha(\beta)),
\end{equation}
where $\alpha(\beta)$ is the surface tension for the interface.
Indeed at large $\beta$ it is easily shown that $\alpha(\beta) = 3/2$.

In the infinite volume system a phase transition takes place at $\beta
= \beta_c$.  In a finite volume, crossover occurs at $\beta_c$, where
many observables change rapidly.  For $\beta < \beta_c$ the ground
state $\left| 0s\right\rangle $ is then identified with the
restored-symmetry-phase (disordered) vacuum $\left| 0\right\rangle $,
and the mixed symmetry state $\left| 0m\right\rangle $ is identified
as the lowest lying mixed-symmetry excitation of the symmetric vacuum.

If the phase transition is first order, all four states coexist.  Thus
to model the rounding of the phase transition, we write the
hamiltonian on the simplified basis of the four unmixed vacuum states,
$\left| 0\right\rangle $, $\left| 1\right\rangle $, $\left|
2\right\rangle $, and $\left| 3\right\rangle $ as follows:
\begin{equation}
\left(
  \begin{array}{cccc}
   \Delta \ \ & \lambda \ \ & \lambda \ \ & \lambda \\ \lambda \ \ & 0
\ \ & \epsilon \ \ & \epsilon \\ \lambda \ \ & \epsilon \ \ & 0 \ \ &
\epsilon \\ \lambda \ \ & \epsilon \ \ & \epsilon \ \ & 0
  \end{array}
\right).
\end{equation}
The row and column labels are in order $0,1,2,3$.  The parameter
$\Delta = \alpha(\beta - \beta_c)$ with $\alpha > 0$ is the energy
difference between the disordered (symmetric) phase state $\left|
0\right\rangle $ and the degenerate ordered phase (broken symmetry)
states $\left| 1\right\rangle $, $\left| 2\right\rangle $, $\left|
3\right\rangle $.  This difference vanishes at crossover.  The
parameter $\lambda$ gives the mixing strength between the disordered
phase vacuum $\left| 0\right\rangle $ and the degenerate ordered phase
vacuums, and the parameter $\epsilon$ gives the direct mixing strength
between the ordered vacuums.  The eigenenergies are
\begin{eqnarray}
  E_{0s} &=& \epsilon + {1\over 2} \Delta - {1\over 2} \sqrt{\Delta^2
- 4\epsilon\Delta + 4\epsilon^2 + 12\lambda^2} \nonumber \\ E_{0m} &=&
-\epsilon \label{eq:toylevels} \\ E_{1s} &=& \epsilon + {1\over 2}
\Delta + {1\over 2} \sqrt{\Delta^2 - 4\epsilon\Delta + 4\epsilon^2 +
12\lambda^2}. \nonumber
\end{eqnarray}
As expected, there are two states belonging to the symmetric
representation of $S_3$: one, the ground state $\left| 0s\right\rangle
$ and the other, an excited state $\left| 1s\right\rangle $; and there
is one twofold degenerate mixed symmetry state $\left| 0m\right\rangle
$ as before.  For large $\beta$ we expect the state $\left|
1s\right\rangle $ to become degenerate with a new excited mixed
symmetry state $\left| 1m\right\rangle $ (not included in our
four-component model), in the same pattern as the $\left|
0s\right\rangle $ and $\left| 0m\right\rangle $ states.  With our
criteria for assigning transverse planes to the four phases (see
Sec.~\ref{sec:cyl} below), the numerical simulation indicates that
$\epsilon << \lambda$.  Thus in the region $\epsilon << |\Delta|$, we
may approximate the energy levels with
\begin{eqnarray}
  E_{0s} &=& {1\over 2} \Delta - {1\over 2} \sqrt{\Delta^2 +
12\lambda^2} \label{eq:E0s} \\ E_{0m} &=& 0 \label{eq:E0m} \\ E_{1s}
&=& {1\over 2} \Delta + {1\over 2} \sqrt{\Delta^2 + 12\lambda^2}.
\label{eq:E1s}
\end{eqnarray}

Figure \ref{fig:toya} summarizes the expected behavior of the
excitation energies, as a function of $\alpha(\beta -
\beta_c)/\lambda$, relative to the ground state energy, which has been
renormalized to zero in this figure.  The lowest two energy
differences in this figure come from Eqs.\ (\ref{eq:E0s}-\ref{eq:E1s})
and the energy difference $E_{1m}$ is simply a sketch.  We see that in
a finite system the crossover results in a smooth connection between
the energy levels on either side of $\beta_c$.  In the infinite volume
limit the crossover is much more rapid and the upper wings of the
curves in Fig.~\ref{fig:toya} level off as a consequence of higher
level crossings.

By introducing a four-level system we have assumed a first order phase
transition.  In a continuous phase transition the states $\left|
0\right\rangle $ and $\left| 0s\right\rangle $ should be equivalent,
and a three-level system would suffice.  Although the model focuses on
the lowest few levels, it can, of course, be enlarged to incorporate
other excited levels as well.  Indeed, in order to incorporate the
excited levels quantitatively, at least eight levels are needed: four
for the states already considered, and four more for the corresponding
excited states.

Let us be more explicit about the excitations of the ordered phase.
In the infinite volume limit these vacuums are not mixed.  We
postulate that the lowest excitation of the state $\left|
1\right\rangle $ is a state $\left| 1*\right\rangle $ of mixed
symmetry, reachable by acting on the vacuum with the zero momentum
spin operator
\begin{equation}
   s(z) = \sum_{x,y} s_{(x,y,z)}.  \label{eq:spinav}
\end{equation}
We shall often refer to the corresponding Schr\"odinger-picture
operator $s \equiv s(0)$.  Without loss of generality we introduce
only one new state as follows:
\begin{equation}
  s\left| 1\right\rangle = \gamma \left| 1\right\rangle + \delta
\left| 1*\right\rangle .
\end{equation}
The corresponding excitations in the other vacuums are reached by a
Z(3) transformation $R$ in analogy with Eq.\ (\ref{eq:Zrotation}) as
follows
\begin{equation}
  \left| 2*\right\rangle = R\left| 1*\right\rangle \ \ \ \left|
3*\right\rangle = R\left| 2*\right\rangle .
\end{equation}
Since $RsR^{-1} = \exp(-2\pi i/3)s$, the Z(3) symmetry requires
\begin{eqnarray}
  s\left| 2\right\rangle &=& e^{2\pi i/3} (\gamma \left|
2\right\rangle + \delta \left| 2*\right\rangle ) \\ s\left|
3\right\rangle &=& e^{-2\pi i/3} (\gamma \left| 3\right\rangle +
\delta \left| 3*\right\rangle ).
\end{eqnarray}
Now just as mixing between the degenerate vacuums leads to a symmetric
state $\left| 0s\right\rangle $ and two mixed symmetry states $\left|
0mk\right\rangle $ ($k=1,2$ labels the two degenerate components), we
expect mixing among the excited state counterparts to result in a
symmetric state $\left| 1s\right\rangle $ and a mixed symmetry state
$\left| 1mk\right\rangle $.  With this notation we are implicitly
identifying these states with the levels of Fig.\ \ref{fig:toya}.
{}From the $S_3$ symmetry we then obtain explicit formulas for the
matrix elements of the spin operator between these states:
\begin{eqnarray}
  \sum_{k=1,2} \left| \left\langle 1s\right|s\left| 0mk\right\rangle
\right|^2 &=& \delta^2 \nonumber \\ \sum_{k=1,2} \left| \left\langle
0s\right|s\left| 0mk\right\rangle \right|^2 &=& \gamma^2 \nonumber \\
\sum_{k=1,2} \left| \left\langle 0s\right|s\left| 1mk\right\rangle
\right|^2 &=& \delta^2 \nonumber \\ \sum_{k=1,2} \left| \left\langle
1s\right|s\left| 1mk\right\rangle \right|^2 &=& \gamma^2.
\label{eq:Mme}
\end{eqnarray}
We note also that if mixing among the degenerate states is weak the
symmetric operator $|s|^2$ satisfies
\begin{equation}
  \left| \left\langle 1s\right|\,|s|^2\,\left| 0s\right\rangle
\right|^2 = \sum_{k=1,2} \left| \left\langle 1mk\right|\,|s|^2\,\left|
0mk\right\rangle \right|^2.  \label{eq:Sme}
\end{equation}

\subsection{Correlations}

We now write working formulas for the correlations between the spin
operators $s$ and $|s|^2$, based on the lowest lying levels discussed
above.

First, we observe that matrix elements must obey selection rules of
the $S_3$ symmetry.  For example, the spin operator $s$ is of mixed
symmetry.  The two components of the operator are just $(\Re s, \Im
s)$.  Thus matrix elements $\left\langle n\right|s\left|
m\right\rangle $ vanish if $\left| m\right\rangle $ and $\left|
n\right\rangle $ are both symmetric.  Since $|s|^2$ is a symmetric
operator, matrix elements $\left\langle n\right||s|^2\left|
m\right\rangle $ vanish if $\left| m\right\rangle $ and $\left|
n\right\rangle $ are not in the same $S_3$ representation.

Therefore, the leading nonvanishing terms in the correlation for
$|s|^2$ from Eq.\ (\ref{eq:corr}) are
\begin{eqnarray}
  \left\langle |s(z)|^2 |s(0)|^2\right\rangle &=& \left|\left\langle
0s\right|\,|s|^2\,\left| 0s\right\rangle \right|^2 +
\left|\left\langle 1s\right|\,|s|^2\,\left| 0s\right\rangle \right|^2
[e^{-E_{1s}z} + e^{-E_{1s}(L-z)}] \nonumber \\ &+& \left|\left\langle
1m\right|\,|s|^2\,\left| 0m\right\rangle \right|^2
[e^{-(E_{1m}-E_{0m})z} + e^{-(E_{1m}-E_{0m})(L-z)}]e^{-E_{0m}L},
\label{eq:Scorr}
\end{eqnarray}
and for $s$ it is
\begin{eqnarray}
  \left\langle s(z) s^*(0)\right\rangle &= & |\left\langle
0m\right|s\left| 0s\right\rangle |^2 [e^{-E_{0m}z} + e^{-E_{0m}(L-z)}]
+ |\left\langle 1m\right|s\left| 0s\right\rangle |^2 [e^{-E_{1m}z} +
e^{-E_{1m}(L-z)}] \nonumber \\ &+& |\left\langle 1s\right|s\left|
0m\right\rangle |^2 [e^{-(E_{1s}-E_{0m})z} +
e^{-(E_{1s}-E_{0m})(L-z)}]e^{-E_{0m}L} \\ &+& |\left\langle
1m\right|s\left| 0m\right\rangle |^2 [e^{-(E_{1m}-E_{0m})z} +
e^{-(E_{1m}-E_{0m})(L-z)}]e^{-E_{0m}L}, \label{eq:Mcorr}
\end{eqnarray}
The six transitions taken into account in these expressions are
indicated in Fig.\ \ref{fig:toya}.

Clearly, in order for the several transitions for be discernible in
the fits to correlation functions, it is necessary that the spectral
components be both strong and well-separated.  With our data we are
able to distinguish two spectral components for the mixed operator and
find only one significant spectral component for the symmetric
operator.  These transitions correspond to dropping all terms with the
factor $\exp(-E_{0m}L)$.  For the mixed operator, this approximation
can be justified as follows: (1) Over the range $\beta < \beta_{c}$,
the factor $\exp(-E_{0m}L)$ is small (at the largest, of order 1/10).
This factor multiplies the third term on the rhs in both expressions
and the fourth term in the second expression.  Appealing to our
phenomenological model and Eqs.\ (\ref{eq:Mme}) and (\ref{eq:Sme}),
which makes it possible to compare the second and third terms, we see
that we may drop terms in $\exp(-E_{0m}L)$ from the summation, thereby
eliminating all transitions leading to the level $E_{0m}$.  (2) Over
the range $\beta > \beta_{c}$, the spectral lines for all transitions
from the levels $1s$ and $1m$ to $0s$ and $0m$ are too close to be
resolvable with our statistics.  Thus the second spectral component in
the mixed operator is presumably a composite of all three transitions
permitted by the selection rules.  For the symmetric operator our
failure to locate the second spectral component presumably reflects an
insufficiently strong coupling to this operator.

We are left, finally, with a three-parameter expression for the
symmetric operator and a four-parameter expression for the mixed
operator:
\begin{eqnarray}
  \left\langle |s(z)|^2 |s(0)|^2|\right\rangle &=& A_{1} + A_{2}
[e^{-E_{1s}z} + e^{-E_{1s}(L-z)}] \nonumber \\ \left\langle s(z)
s^*(0)\right\rangle &= & C_{1} [e^{-E_{0m}z} + e^{-E_{0m}(L-z)}] +
C_{2} [e^{-E_{1m}z} + e^{-E_{1m}(L-z)}].  \label{eq:fit}
\end{eqnarray}

\section{Numerical Simulation} \label{sec:NumSim}

Since the simulation of tunneling effects requires an algorithm that
gives efficient and rapid sampling of the phase space, particularly in
the crossover region, we used the Swendson-Wang (SW)
algorithm\cite{ref:SW}. The code was checked by comparing measured
observables in selected extensive runs with results from simulations
using a heat bath algorithm.

Simulations were carried out on cubic lattices of size $N^3$, for $N =
16$, 20, 24, 30, 48, and 64 in order to confirm the first order nature
of the phase transition through the use of finite-size scaling
analysis.  For this purpose we measured the specific heat and the
fourth-order cumulant.  Table \ref{tab:data_cubic} summarizes the
extent of this data sample.  The majority of the simulations were
carried out on ``cylindrical'' lattices to obtain the spectrum of the
transfer matrix.  These lattices were of size $120\times 20^2$ for 18
values of $\beta$ and with size $120\times 30^2$ for 17 values of
$\beta$, as summarized in Table \ref{tab:data_cyl}.  Runs of as many
as two million sweeps made it possible to gain control of the
correlated data and to obtain good estimates of the parameter errors.

\subsection{Simulations on cubic lattices}

In Fig.~\ref{fig:hen64} is shown a histogram in the average energy for
the $64^3$ lattice near $\beta_c$.  A clean separation of the phases
is apparent.  The importance of finite size scaling for determining
the order of a phase transition has been repeatedly
emphasized\cite{ref:Karsch,ref:Fukugita}. Although these references
already provide excellent confirmation of the first order character of
the phase transition, our high statistics results make an even
stronger case.  Indeed, with such a clean separation of phases in the
$64^3$ lattice, it is scarcely necessary to belabor the point.
Nevertheless, we present the finite size scaling results for the sake
of completeness.  The fourth order cumulant\cite{ref:CLB}, %
\begin{equation}
  V_L(\beta,V) = 1 - \frac{\left\langle E^4\right\rangle}{3
\left\langle E^2\right\rangle^2},
\end{equation}
is minimum near the crossover $\beta_c$.  At a continuous phase
transition, the minimum value $V_L(\beta_c,V)$ tends to $\frac{2}{3}$
in the infinite volume limit.  At a first order phase transition,
however, the limit is not so constrained, but obeys a scaling law,
\begin{equation}
  V_L(V)_{min} \rightarrow 1 - \frac{(E_+^2 + E_-^2)^2}{12(E_+E_-)^2}
+ O(1/V),
\end{equation}
where $E_+$ and $E_-$ are the most probable energies in the two
coexisting phases.

The minimum value of the cumulant was found by combining measurements
at a range of $\beta$ values near $\beta_c$, using a
Ferrenberg-Swendson ``scanning'' or ``histogram''
technique\cite{ref:FS}.  The resulting minimum values are plotted in
Fig.~\ref{fig:scal_cm}.  A linear fit yields the asymptotic value
0.647(3), clearly distinct from $\frac{2}{3}$.

We turn now to the specific heat.  At finite volume the specific heat
$C_v$ peaks at the crossover.  The maximum value $C_{v,max}$ increases
with increasing volume.  If the phase transition is first order, the
peak scales as
\begin{equation}
  C_{v,max} = a + bV.
\end{equation}
The maximum is again determined using the Ferrenberg-Swendson
technique.  These values are plotted in Fig.~\ref{fig:scal_cv}.  A
linear increase is apparent.

A value of $\beta_c$ can be fixed, either from the peak in the
specific heat, or from the minimum of the cumulant $V_L$.  The two
values do not necessarily agree at finite volume, but should agree in
the infinite volume limit.  In the infinite volume limit we find
$\beta_{c,\infty} = 0.36704(2)$ in agreement with Gavai, Karsch and
Petersson\cite{ref:Karsch}.

\subsection{Simulations on cylindrical lattices} \label{sec:cyl}

Measurements on the asymmetrical lattices were taken every 250 SW
sweeps.  Observables recorded were these: the spin averages $s(z)$ as
a function of $z$ [Eq.\ (\ref{eq:spinav})], the average energy,
\begin{equation}
  E = \beta \sum_{(i\mu)}\Re s^*_i s_{i+\mu}/V,
\end{equation}
and the number of clusters $N_c$.  Subsequent analysis produced the
symmetric and mixed operator correlations and spectrum, the mean
spins, and the projected-spin order parameter
\begin{equation}
   s_{proj} = \max \Re(\bar s, e^{2\pi i/3}\bar s, e^{-2 \pi i/3} \bar
s) ,
\end{equation}
where $\bar s = \sum s_i/V$.  Also constructed were interface
statistics.  They are described below.  For observables not discussed
here, see Ref.~\cite{ref:wang}.

\paragraph{Spectrum of the Transfer Matrix} Correlations in the
operators $|s(z)|^2$ and $s(z)$ were measured and fit to the formulas
(\ref{eq:fit}) for both transverse sizes $20^2$ and $30^2$.  As usual,
fluctuations in the measurements were strongly correlated in $z$, so
it was necessary to determine these correlations and incorporate them
in the $\chi^2$ analysis.  Because of the large size of the data
sample, it was possible to use all principal factors in the analysis
of covariance.  The spectrum was determined from a global fit to the
data.  The fitting range began at a minimum distance $z_{min}$ and
extended to the full length of the lattice.  The minimum distance was
varied until a semblance of a plateau in the spectrum was reached,
within the determined statistical errors.  The values quoted are based
on the minimum distance that gave the highest confidence level for the
fit.  The minimum distance thus determined varied smoothly from 2 for
the mixed operator at the smallest $\beta$ where the correlation
length is shortest to 8 at the largest $\beta$ where the correlation
length is longest.  For the symmetric operator $z_{min}$ was 5 for the
smallest and largest $\beta$'s, ranging gradually to $15-20$ for the
intermediate values where the correlation length is longest.  The
resulting spectrum is summarized in Figs.~\ref{fig:mass20_sym},
\ref{fig:mass30_sym}, \ref{fig:mass20_mix}, and \ref{fig:mass30_mix}
and in Tables~\ref{tab:20} and \ref{tab:30}.  There is obviously a
strong resemblance with features of the simple model of
Sec.~\ref{sec:model}.  From Fig.~\ref{fig:mass20_mix} and
\ref{fig:mass30_mix}, we see that the correlation length tends to
infinity as $\beta$ increases above $\beta_c$, just as with lattices
of similar geometry in SU(3) Yang-Mills theory\cite{ref:APE}.  As we
have seen in the simple model, this feature is an expected consequence
of finite volume tunneling between the degenerate ordered phases.

\paragraph{Tunneling Statistics} In Fig.~\ref{fig:tunnel} we plot the
slice spin averages $s(z)$ for a representative configuration at
$\beta > \beta_c$, showing tunneling between the ordered phases.  We
devised two statistics: $N_{do}$, to give a measure of the number of
phase boundaries between the disordered and one of the ordered
domains, and a statistic $N_{oo}$, for phase boundaries between two
ordered domains.  A portion of the lattice was considered to be in the
disordered phase on the plane $z$, if $|s(z)| < 0.23$ for at least
three consecutive values of $z$.  If $|s(z)| > 0.23$ for at least
three consecutive values of $z$, the lattice plane was considered to
be in one of the three ordered phases, according to the value of $\arg
s(z)$.  The value 0.23 was chosen to correspond to the minimum of the
histogram of occurrences of values of $|s|$ at the crossover, and so
corresponds to a value intermediate between the disordered and ordered
phases.  The requirement of three consecutive planes was adopted to
permit occasional excursions from the ideal value of 0.23 within a
single phase.  We encountered no configurations that did not have at
least three consecutive planes.  Indeed the phase coherence was
extremely high with many lattices consisting of a single phase.

Obviously our classification criteria are arbitrary.  Our approach
differs from that of Karsch and Patk\'os\cite{ref:KP} who classified
all boundaries as type oo for $\beta > \beta_c$.  The ambiguity all
methods must deal with is distinguishing a broad interface between two
ordered phases from a transition to an intermediate disordered phase.
Any definition must recognize, however, that in perturbation theory,
finite volume mixing necessarily produces a disordered phase
intermediate for $\beta$ slightly above $\beta_c$.

Shown in Figs.~\ref{fig:oo_20}, \ref{fig:n20}, and \ref{fig:n30} are
results for the measure of the mean numbers $N_{oo}$ and $N_{do}$.  It
is apparent that with our definitions of these boundaries, a
transition between two ordered phases is unlikely to take place
directly, but proceeds through what we identify as a disordered phase
intermediate.  (Figure~\ref{fig:oo_20} shows the number of $oo$ phase
boundaries for the $120\times 20^{2}$ lattices.  The corresponding
number for $120\times 30^{2}$ is negligible.)  Thus if we interpret
these results in terms of the model of Sec.~\ref{sec:model}, we find
that $\epsilon << \lambda$.  Thus the approximation,
Eqs.~(\ref{eq:E0s}-\ref{eq:E1s}) applies.  Inverting these
expressions, the parameters $\Delta$ and $\lambda$ of the simple model
can then be derived from the observed spectrum:
\begin{eqnarray}
   \Delta &=& E_{1s} - 2 E_{0m} \\ \lambda &=& \sqrt{E_{0m}(E_{1s} -
E_{0m})/3}
\end{eqnarray}

Let us estimate the number of ``do'' (ordered/disordered) phase
boundaries expected in our four-component model.  Since $\lambda$ in
the four-component hamiltonian mixes the disordered and ordered
states, we can introduce a chemical potential for $do$ boundaries by
replacing $\lambda$ with $\lambda e^{\mu}$.  Thus the number of $do$
boundaries is just
\begin{equation}
  \left\langle N_{do}\right\rangle = \lambda\frac{\partial}{\partial
\lambda} \ln Z,
\end{equation}
but from Eqs.~(\ref{eq:part_func}), (\ref{eq:defT}), and
(\ref{eq:E0s}) we have $\ln Z \approx -L E_{0s}(\lambda)$, so
\begin{equation}
    \left\langle N_{do}\right\rangle \approx
6L\lambda^2/\sqrt{\Delta^2+12\lambda^2}.
\end{equation}
This predicted value (for $\beta > \beta_{c}$) is plotted together
with the observed number in Fig.\ref{fig:n20}.  The agreement is quite
satisfactory.

\paragraph{Surface Tension} The surface tension between two ordered
phases can be estimated in perturbation theory (dilute interface
approximation) through
\begin{equation}
  \lambda^2/\Delta = e^{-\beta\alpha_{ord}A}.
\end{equation}
The quantity on the left is the transition probability between two
ordered phases via a disordered phase intermediate in second order
perturbation theory, based on the four-component hamiltonian, and the
quantity on the right is the Boltzmann weight for the interface.  Thus
estimated, it is plotted in Fig.~\ref{fig:surf_ten}.  There is a strong
dependence on the transverse size, suggesting a significant finite
size correction.  Karsch and Patk\'os\cite{ref:KP} measured the
surface tension in the Potts model and also find a strong dependence
on transverse size.  They follow a statistical approach based on the
frequency of $oo$ interfaces, using a different definition from ours,
which excludes all $do$ interfaces.  For example the configuration of
Fig.~\ref{fig:tunnel} yields nine ``bare domain walls'' in their
language, but all presumably classified in their scheme as ``defects''
and not genuine interfaces.  Perhaps it is not surprising, therefore,
that our surface tension is two to three times lower than theirs
(after making allowances for different conventions in the
hamiltonian).  Clearly the determination of surface tension is model
dependent.

\section{Summary and Discussion}

Our high statistics study of the three-state three-dimensional Potts
model using the Swendson-Wang updating scheme once again confirms the
first order character of the phase transition.  At volumes as large as
$64^3$ the separation of coexisting phases is so clear that a
sophisticated finite size scaling analysis is scarcely necessary.

Exploiting an $S_3$ symmetry in lattices with one long dimension, we
have obtained the lowest spectral levels of the transfer matrix in
this model and find excellent agreement with the spectrum of a simple
four-component model, featuring a first order phase transition.  This
analysis provides a clear explanation of the mechanism that gives rise
to an infinite correlation length in the low temperature phase.

The statistics of phase boundaries at low temperature are consistent
with a perturbative treatment of tunneling in the four-component
model.  With our assignment of planes to phases, we find that
ordered-ordered phase boundaries almost always involve an intermediate
disordered phase.  We introduced a spectral method for estimating the
surface tension, based on the four-component model.

One important goal of finite size spectral analysis is to remove
tunneling-related finite-size effects from the spectrum, with the hope
of extracting the infinite volume values of the excitation spectrum.
Would this be feasible using our methods?  Unfortunately, to remove
significant finite size effects apparently requires introducing more
parameters into the four-component model and into our fitting
functions than data of the quality of ours warrants.  The four
component model would have to be augmented by at least four more
components.  Thus we must rely upon alternate, empirical methods.  For
example, Fukugita et al\cite{ref:Fukugita} calculate in a cubic
volume.  In the vicinity of the phase transition, they classify
configurations into two groups: disordered and ordered, according to
the value of the global order parameter.  They then measure
``pure-phase'' correlation lengths in each subset.  For small volumes
there is a region of overlap in which this classification risks
misidentification of the phase.  The contamination of incorrectly
classified configurations decreases as the volume is increased and the
overlap decreases.  Thus one may hope for an empirically determined
extrapolation to the infinite volume limit.

\acknowledgements

Computations were carried out in the initial portion of this work on
an IBM 3090/600S at the Utah Supercomputing Institute and on a Cray
Y/MP at the San Diego Supercomputer Center, and in later portion, on
IBM RS 6000/320 workstations in the Department of Physics, University
of Utah.

%
\figure{ Phenomenological model of the lowest three energy levels at
crossover.  The ground state energy has been adjusted to zero at all
$\beta$.  The level $E_{1m}$ does not come from the model; it is based
merely on a guess.  The others come from the four-component model Eq.\
(\ref{eq:E0s}-\ref{eq:E1s}).  Transitions induced by a mixed operator
are indicated with ``M'' and a symmetric operator with ``S''.
\label{fig:toya} }
\figure{ Histogram in the average energy at $\beta = 0.36705$ (near
the critical value) on a $64^3$ lattice, showing cleanly separated
coexisting phases.  \label{fig:hen64}}
\figure{ Finite size scaling of the minimum in the fourth order
cumulant.  \label{fig:scal_cm}}
\figure{ Finite size scaling of the peak in the specific heat.
\label{fig:scal_cv}}
\figure{ Energy level $E_{1s}$ of symmetry $S$ for the $20^2\times
120$ lattice, as a function of $\beta$.  \label{fig:mass20_sym}}
\figure{ The same, but for the $30^2\times 120$ lattice
\label{fig:mass30_sym}}
\figure{ Energy levels $E_{0m}$ and $E_{1m}$ of symmetry $M$ for the
$20^2\times 120$ lattice, as a function of $\beta$.
\label{fig:mass20_mix}}
\figure{ The same, but for the $30^2\times 120$ lattice
\label{fig:mass30_mix}}
\figure{ Plot of the complex modulus and argument of the spin vector
$s(z)$ vs $z$ (averaged over the transverse plane) for a typical
configuration selected from the data sample at $\beta = 0.3672$ (near
the phase transition) on a $20^2 \times 120$ lattice.  The plot symbol
indicates the phase to which the lattice plane is assigned, based on
modulus and argument.  The vertical bars indicate an assigned phase
boundary, based on our arbitrary rule that at least three consecutive
planes must be classified in that phase.  The horizontal line in the
modulus plot indicates our division between ordered and disordered
phases.  Two ordered-ordered phase boundaries appear in this
configuration, each with a disordered phase intermediate.
\label{fig:tunnel} }
\figure{ Average number of phase boundaries separating ordered
symmetry phases for $120\times 20^{2}$.  \label{fig:oo_20}}
\figure{ Average number of phase boundaries separating a disordered
phase from an ordered phase for $120\times 20^{2}$.  Crosses are
measured directly.  Octagons are calculated from the spectrum in the
four-component model.  \label{fig:n20}}
\figure{ Same, but for $120\times 20^{3}$.  \label{fig:n30}}
\figure{ Ordered-ordered interface surface tension determined by
spectral methods in the four-component model.  Octagons are for
$20^{2}$; crosses are for $30^{2}$.\label{fig:surf_ten} }
\narrowtext
\begin{table}
\caption{\label{tab:data_cubic} Data sample for cubic lattices }
\begin{tabular}{cll}
  V & $\beta$ &sweeps ($10^6$) \\ \hline $16^3$	& 0.3663 & 0.6	 \\
$20^3$	& 0.3665 & 0.5	 \\ $24^3$	& 0.3667 & 0.7	 \\
$30^3$	& 0.3669 & 0.7	 \\ $48^3$	& 0.3670 & 0.5	 \\ $64^3$ &
0.36705 & 0.5
\end{tabular}
\end{table}
%
\narrowtext
\begin{table}
\caption{Data sample for cylindrical lattices.  (Million sweeps.)
\label{tab:data_cyl}}
  \begin{tabular}{lcc}

 $\beta$& $120 \times 20^2$ 	& $120 \times 30^2$ 	\\ \hline
.3650 & 1.0			& 1.0			\\	 .3655
& ---			& 1.0			\\	 .3660	&
1.0			& 1.0			\\ .3665 & 1.0
& 1.0			\\ .3668 & 2.0			&
1.0			\\ .3669 & 2.0			&
2.0			\\ .36695 & 2.0			&
2.0			\\ .3670 & 2.0			&
2.0			\\ .36705 & 2.0			&
2.0			\\ .3671 & 2.0			&
2.0			\\ .36715 & 2.0			&
2.0			\\ .3672 & 3.0			&
2.0			\\ .36725 & 2.0			&
2.0			\\ .3673 & 1.0			&
1.0			\\ .3674 & 1.0			&
1.0			\\ .3675 & 1.0			&
1.0			\\ .3678 & 1.0			&
---			\\ .3680 & 1.0			&
1.0			\\ .3685 & 1.0			&
---			\\
  \end{tabular}
\end{table}
%
\widetext
\begin{table}
\caption{Effective masses and couplings for the $120 \times 20^2$
lattice \label{tab:20} }
\begin{tabular}{lllllllll}
 $\beta$ &$A_1$ &$A_2$ &$E_{1s}$& $C_1$ &$E_{0m}$ &$C_2$ &$E_{1m}$ \\
& & & & & & & \\
 \hline .3650 & 2.38(1)&0.5(1) &.231(27)&2.9(1)&.167(5) & 0.18(11)& .54(28) \\
 .3660 &3.20(3)&0.94(7)&.168(10)&3.3(9) &.10(2) & 1.0(9) & .20(7) \\
 .3665 &4.36(4)&1.6(2) &.118(9) &5.8(1) &.072(1) & 0.31(4) & .54(11) \\
 .3668& 6.11(6)&2.6(3) &.109(8) &8.0(1) &.048(1) & 0.30(5) & .34(11) \\
 .3669 & 7.17(7)&2.5(2) &.088(6) &9.3(1) &.0397(6) & 0.34(9) & .41(12)\\
 .36695 & 7.8(1) &2.9(4) &.095(11)&10.1(2)&.037(1) & 0.34(1) &.41(18) \\
 .3670 & 8.51(8)&2.7(4) &.088(10)&10.8(1)&.0346(6) & 0.24(6)& .31(16) \\
 .36705 & 9.59(9)&2.8(4) &.084(9) &11.8(1)&.0302(4) &0.26(8) & .38(17) \\
 .3671 & 10.5(1)&3.0(4) &.084(9) &12.4(1)&.0272(4)& 0.35(5) & .28(8) \\
 .36715 & 11.8(1)&2.8(4) &.082(9)&13.2(3)&.0233(6) & 0.6(3) & .12(5) \\
 .3672 & 13.2(1)&2.8(1) &.082(3)&14.1(4)&.0207(6) & 0.5(3) & .11(7) \\
 .36725 & 14.4(1)&2.6(2)&.081(5) &14.5(6)&.0180(7) & 0.8(4) & .10(5) \\
 .3673 & 15.8(2)&2.8(4)&.085(11)&15.4(4)&.0159(7) & 0.8(2) & .14(9) \\
 .3674 & 18.6(2)&1.8(2)&.068(12)&16.2(3)&.0122(5) & 1.8(8) & .21(7) \\
 .3675 & 20.7(2)&1.5(3)&.076(14)&16.9(3)&.0102(5) & 0.9(3) & .18(8) \\
 .3678 & 25.5(1)&1.3(1)&.142(11)&15.6(5)&.0039(6) & 1.3(1) & .14(3) \\
 .3680 &27.2(1)&1.12(5)&.175(9) &15.9(4)&.0029(4) & 1.2(2) & .23(4) \\
 .3685 &30.4(1)&1.03(6)&.256(13)&15.7(3)&.0007(3) & 1.2(1) & .28(3) \\
\end{tabular}
\end{table}
%
\widetext
\begin{table}
\caption{Effective masses and couplings for the $120 \times 30^2$
lattice \label{tab:30} }
\begin{tabular}{lllllllll}
 $\beta$ &$A_1$ &$A_2$ &$E_{1s}$& $C_1$ &$E_{0m}$ &$C_2$ &$E_{1m}$ \\
& & & & & & & \\
 \hline .3650 & 1.02(1)&0.2(1) &.30(5) &1.2(1)&.182(8) & .1(1) & .39(16) \\
 .3655 & 1.11(1)&0.23(6)&.25(3)&1.29(8)&.157(6) & .15(7) & .45(14) \\
 .3660 & 1.23(1)&0.38(8)&.25(3)&1.56(2)&.143(3) & --- & --- \\
 .3665 & 1.48(1)&0.44(9)&.15(2)&1.74(9)&.101(4) & .24(9) & .30(6) \\
 .3668 & 2.03(3)&0.74(6)&.065(7)&2.3(2) &.052(3) & .59(18) & .16(3) \\
 .3669 & 3.21(6)&1.28(5)&.061(4)&3.7(1) &.031(1) & .72(12) & .13(1) \\
 .36695 &4.09(9)&1.61(4)&.044(4) &5.2(1) &.0262(6) & .39(4) & .19(2) \\
 .3670 & 6.3(2) &2.02(7)&.038(5) &7.3(1) &.0185(3) & .41(3) & .20(2) \\
 .36705 & 9.9(3) &2.1(1) &.033(6) &9.9(1) &.0133(2) & .38(3) & .21(3)\\
 .3671 & 14.4(2)&1.6(1) &.039(6) &11.8(1)&.0092(2) & .55(4) &.14(1) \\
 .36715 & 17.4(3)&1.3(2) &.030(9) &12.6(1)&.0064(2) &.65(3) & .16(1) \\
 .3672 & 19.6(1)&0.98(3)&.066(4) &12.5(2)&.0044(2) &.75(4) & .12(1) \\
 .36725 & 20.9(1)&0.85(8)&.085(7) &12.2(2)&.0028(2)& .86(3) & .12(1) \\
 .3673 & 21.6(1)&0.7(1) &.09(1) &12.6(2)&.0027(3)& .73(4) & .13(1) \\
 .3674 & 23.1(1)&0.5(2) &.11(2) &12.3(3)&.0011(4)& .69(4) & .15(2) \\
 .3675 & 24.1(1)&0.5(1) &.13(1) &12.7(2)&.0010(3)& .8(1) & .19(2) \\
 .3680 & 27.8(1)&0.50(7)&.22(2) &14.0(3)&.0001(3) &.50(5) & .22(1) \\
\end{tabular}
\end{table}
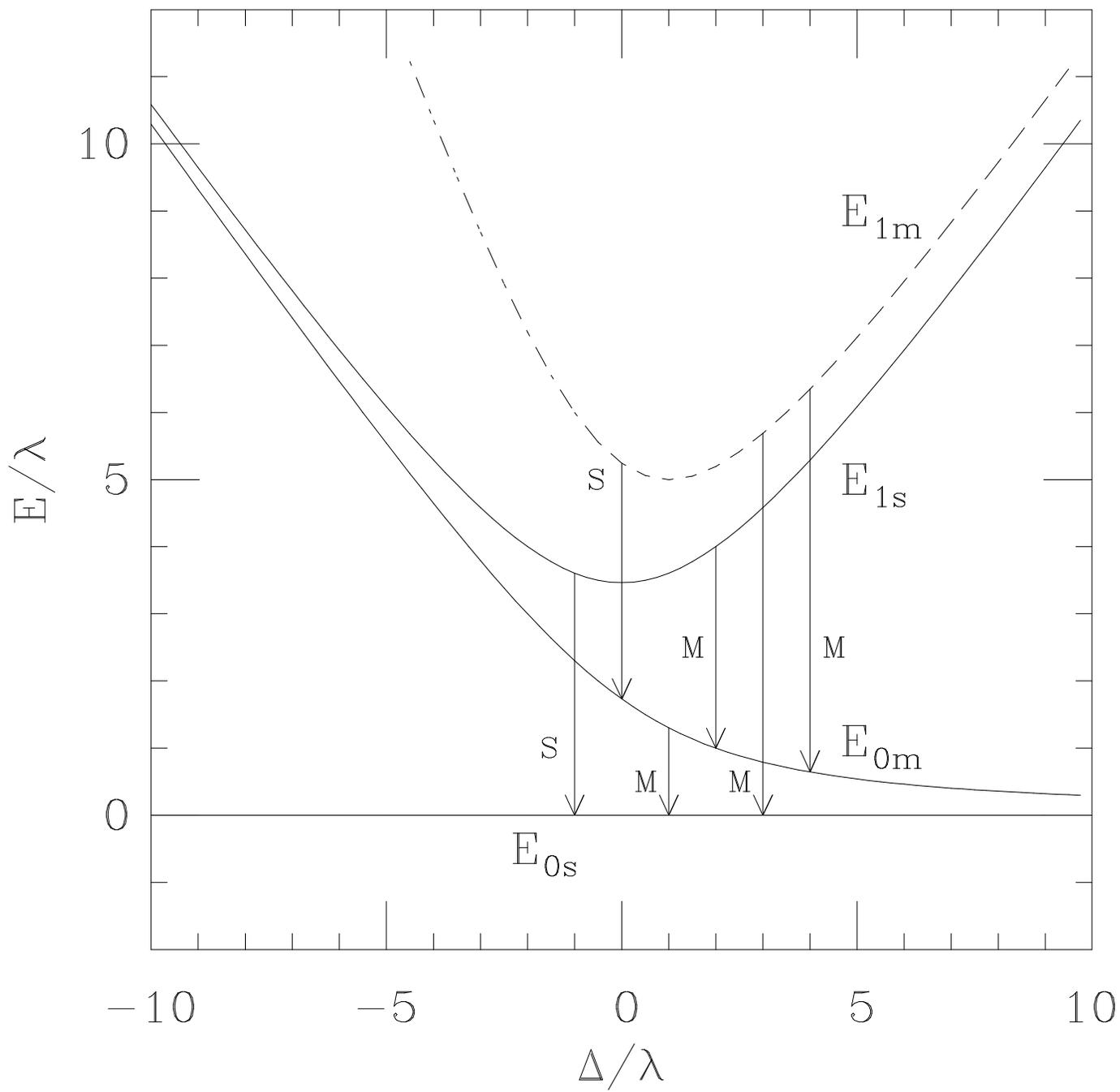
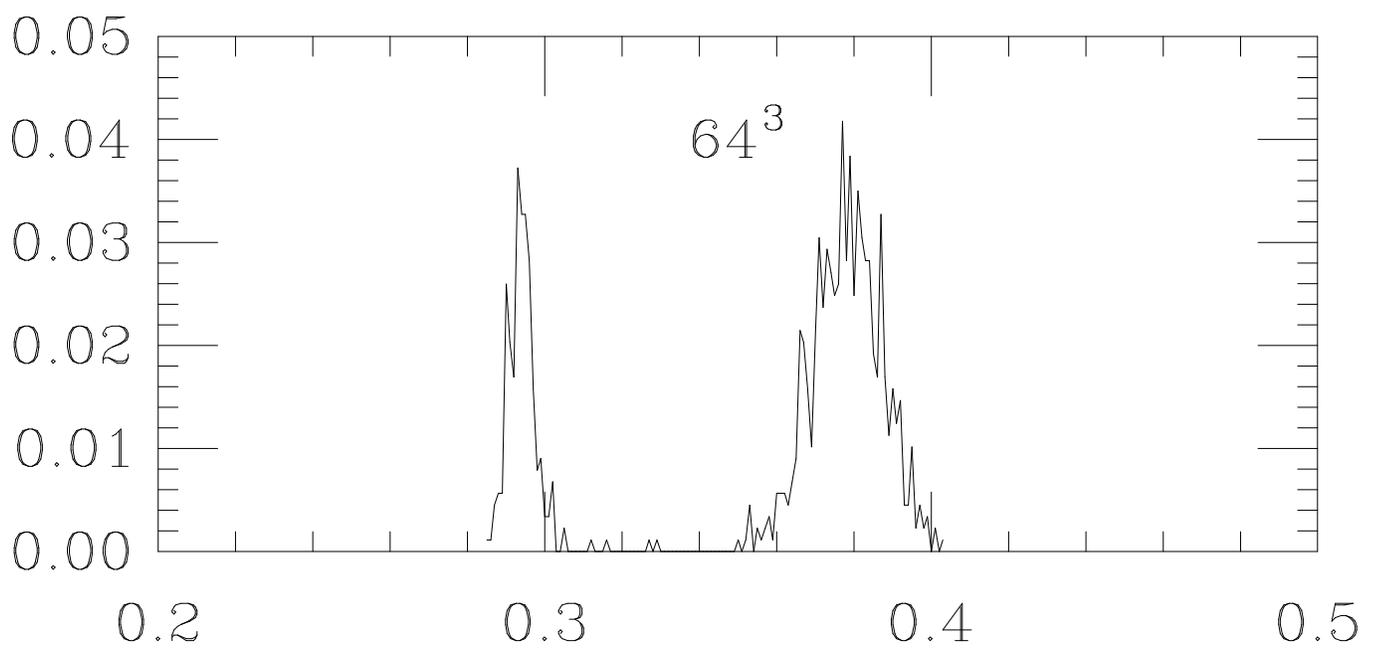
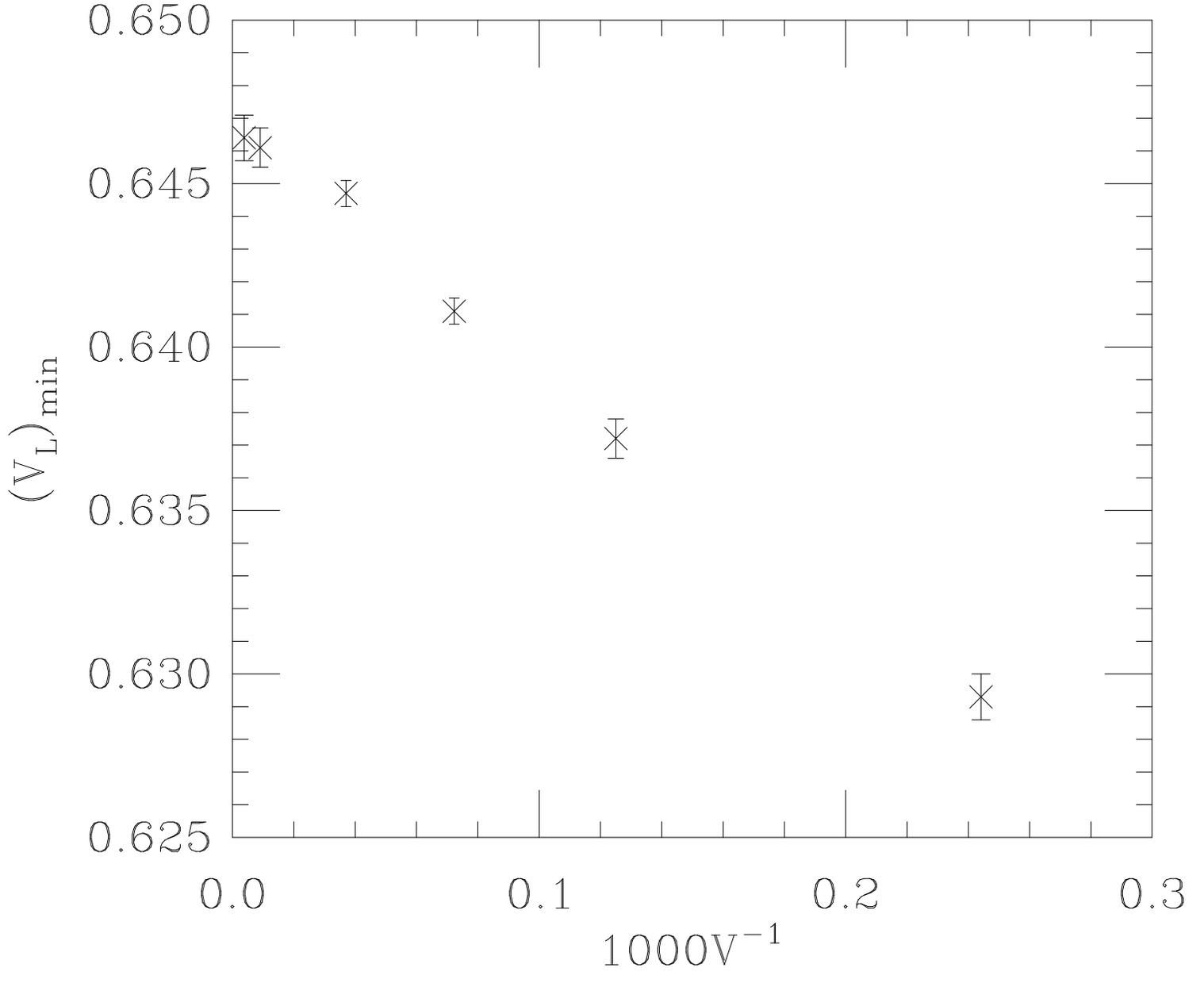
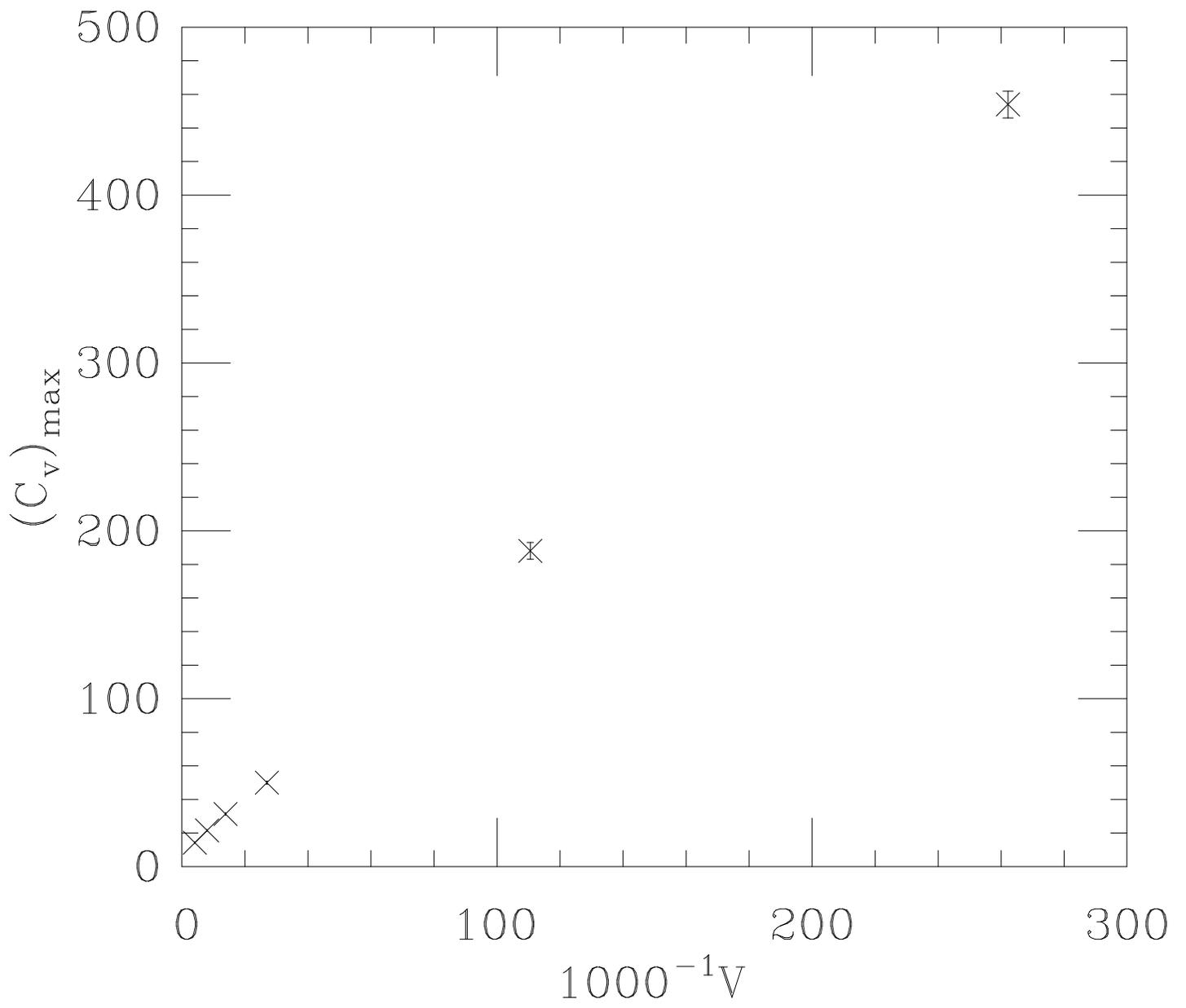
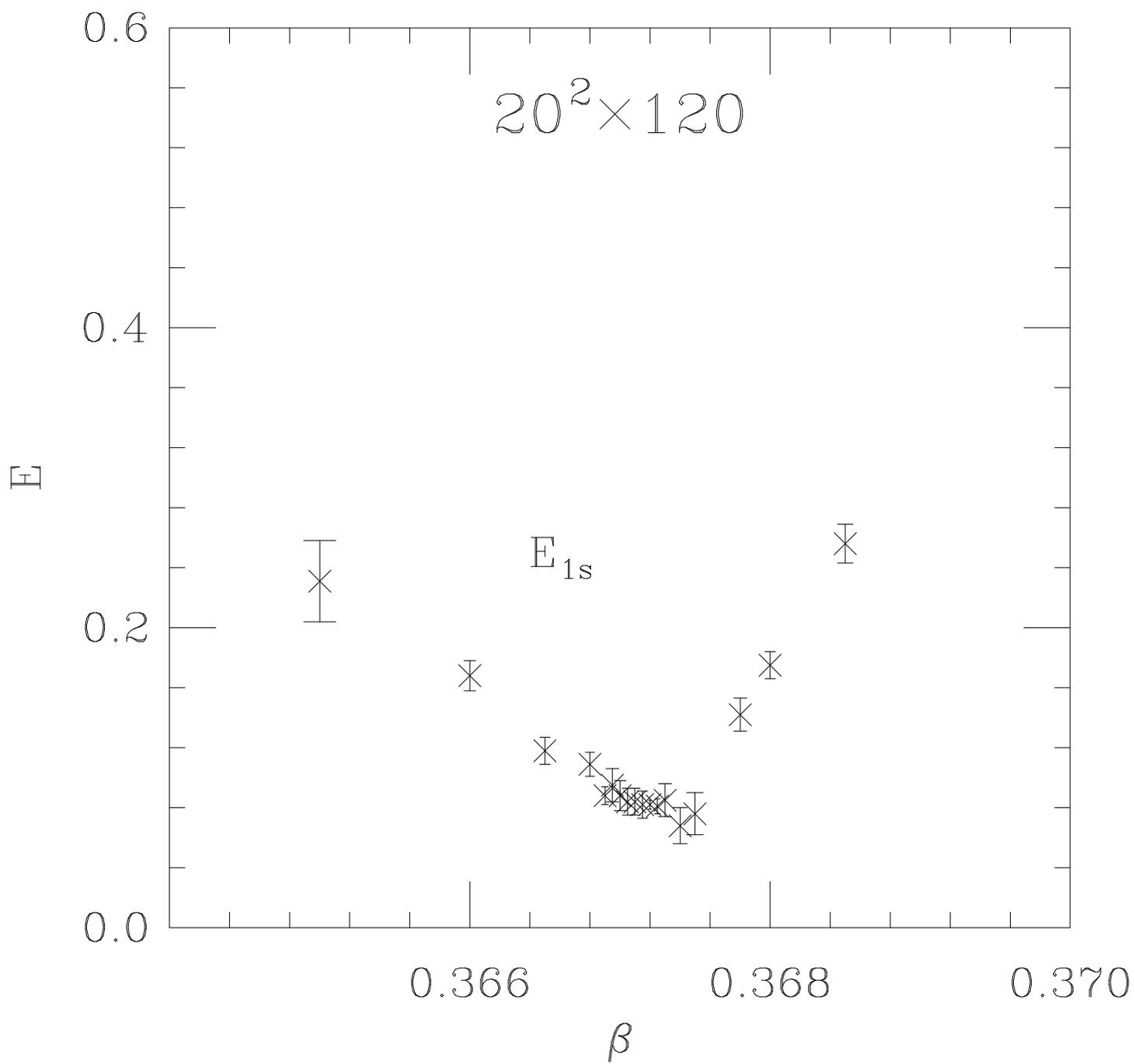
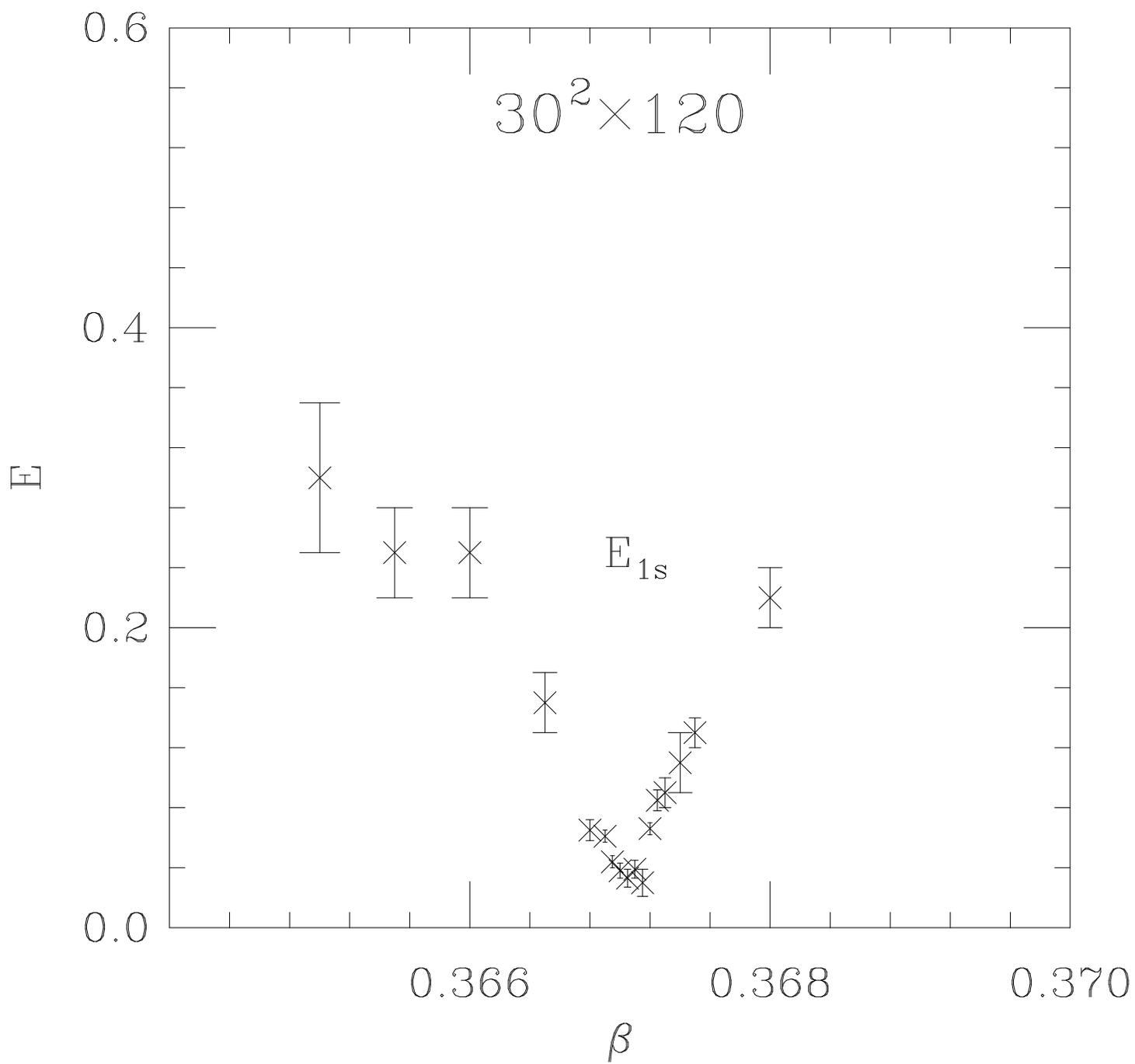
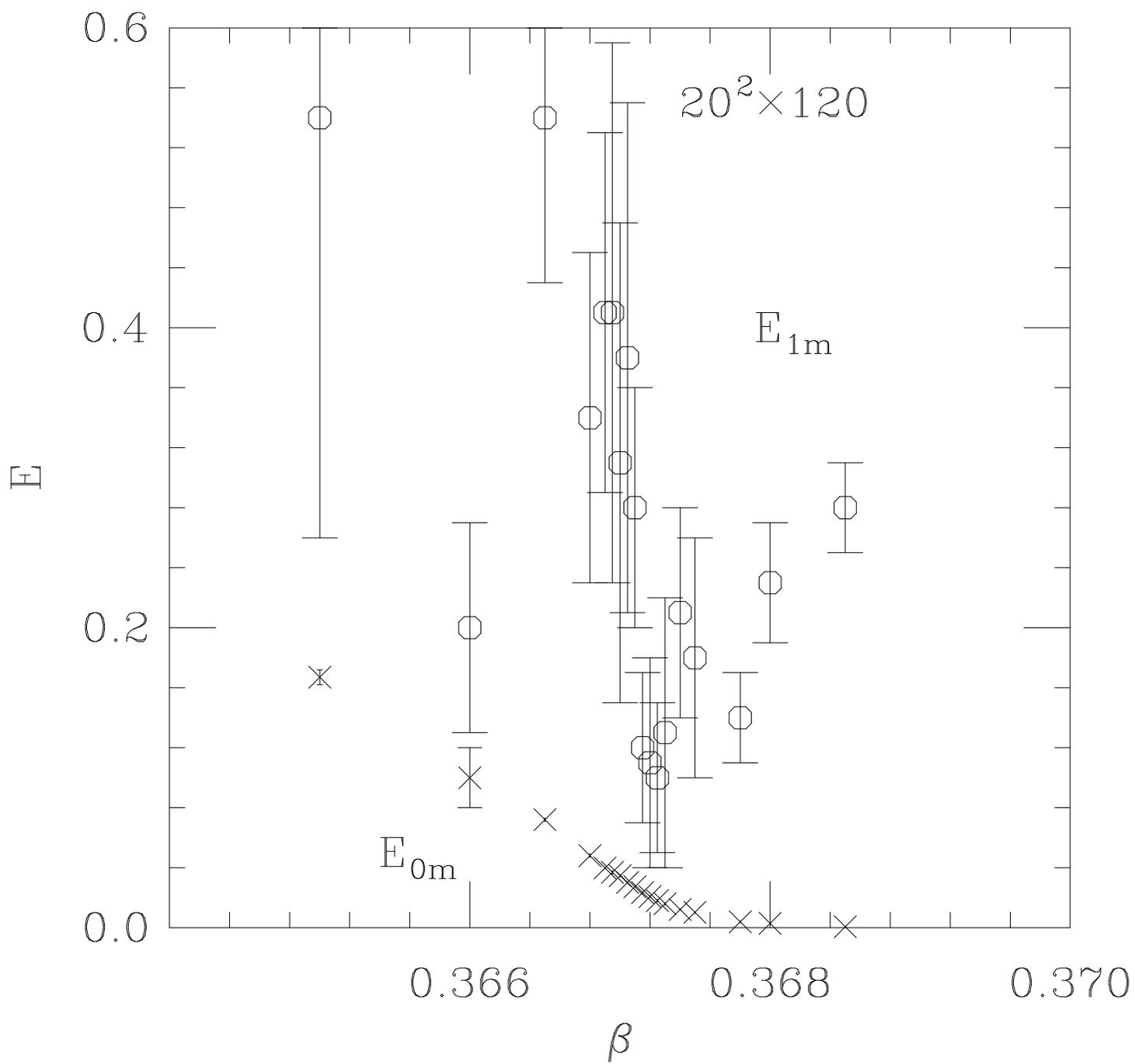
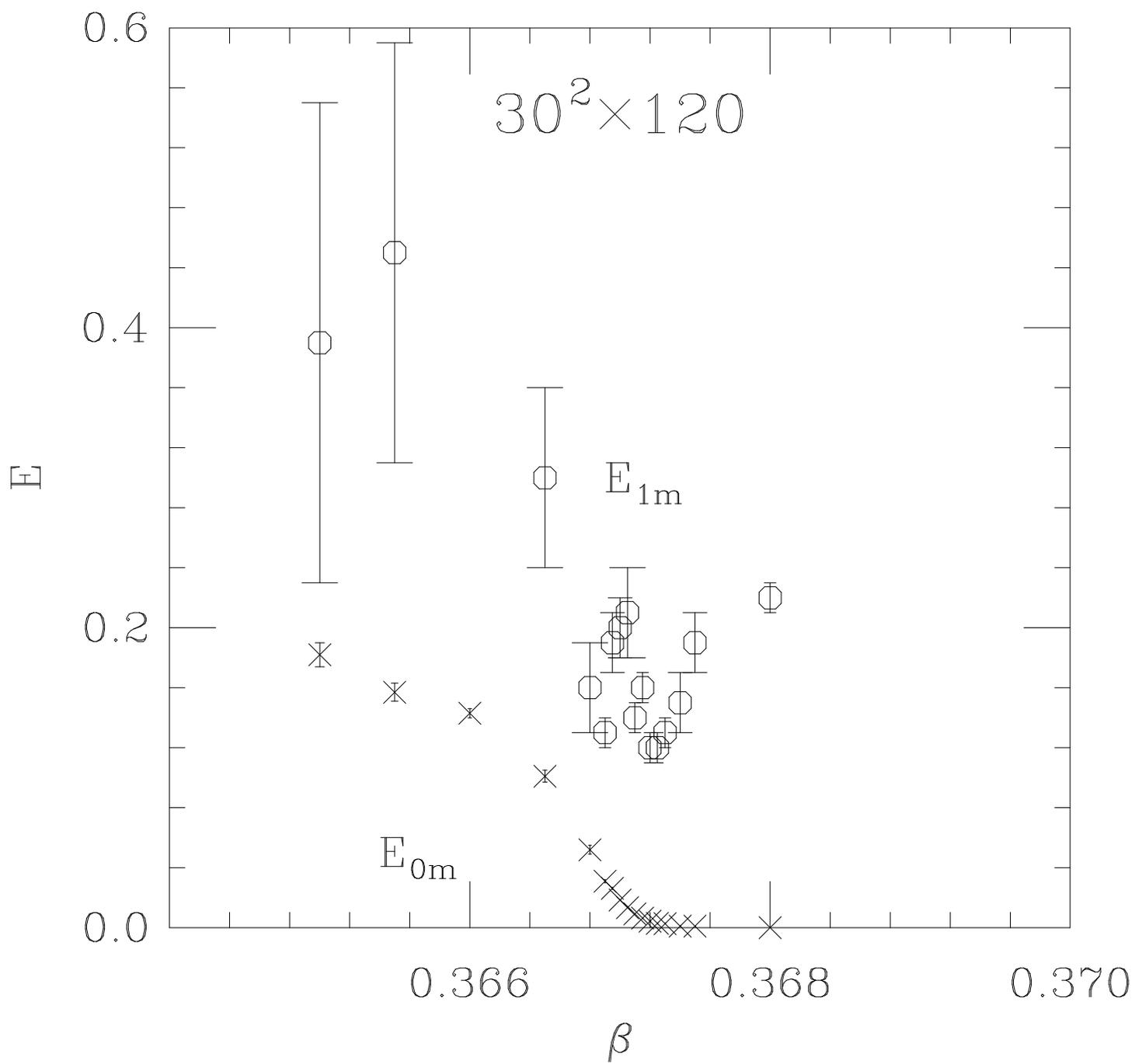
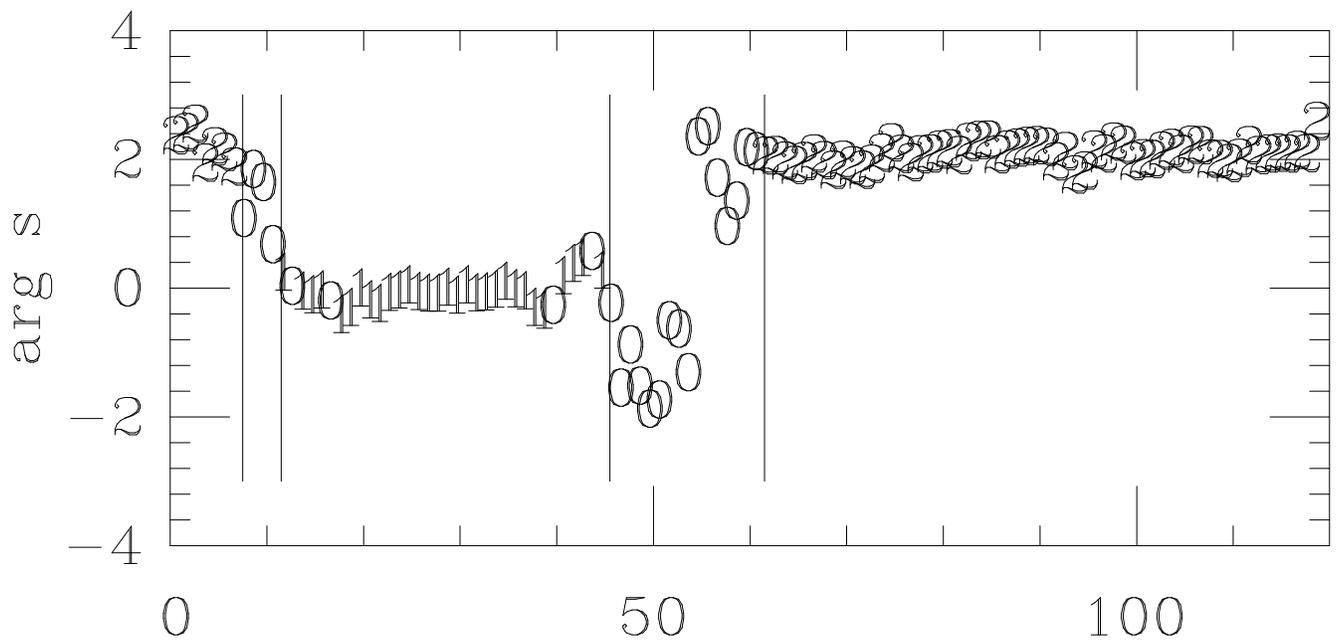
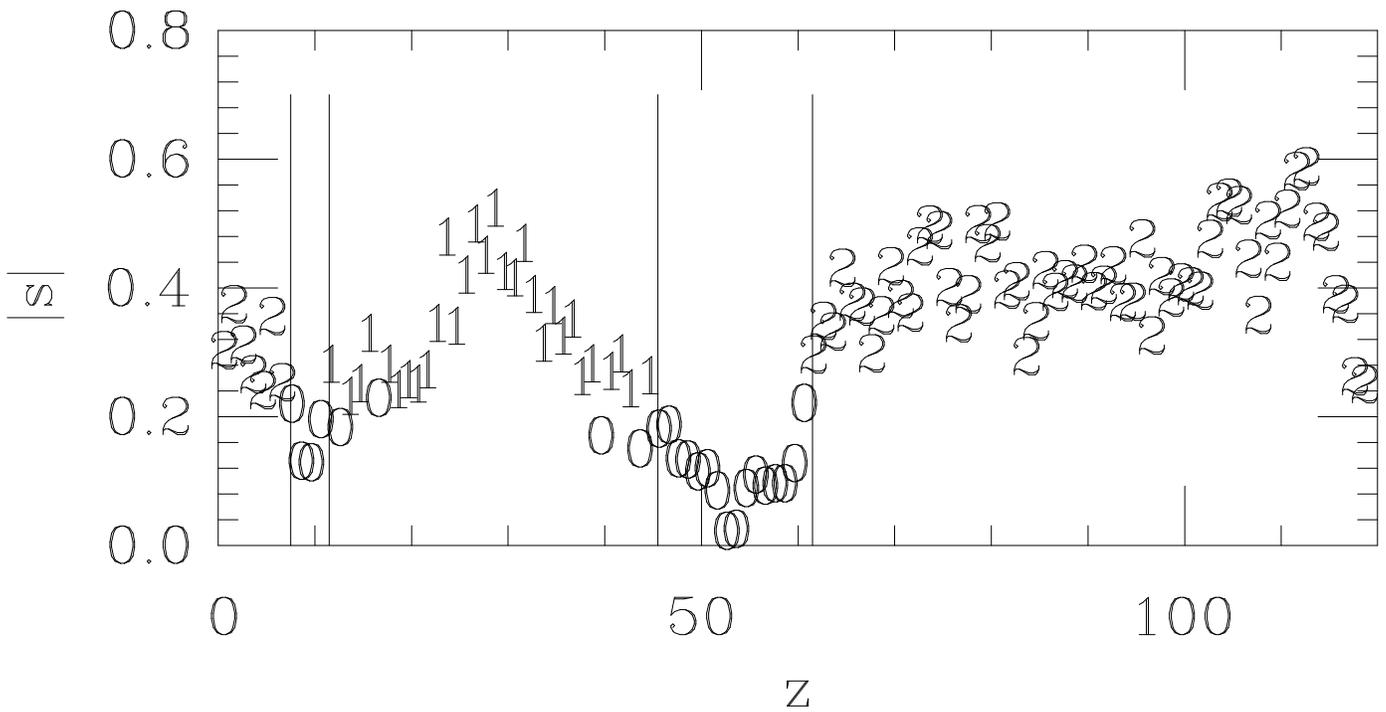
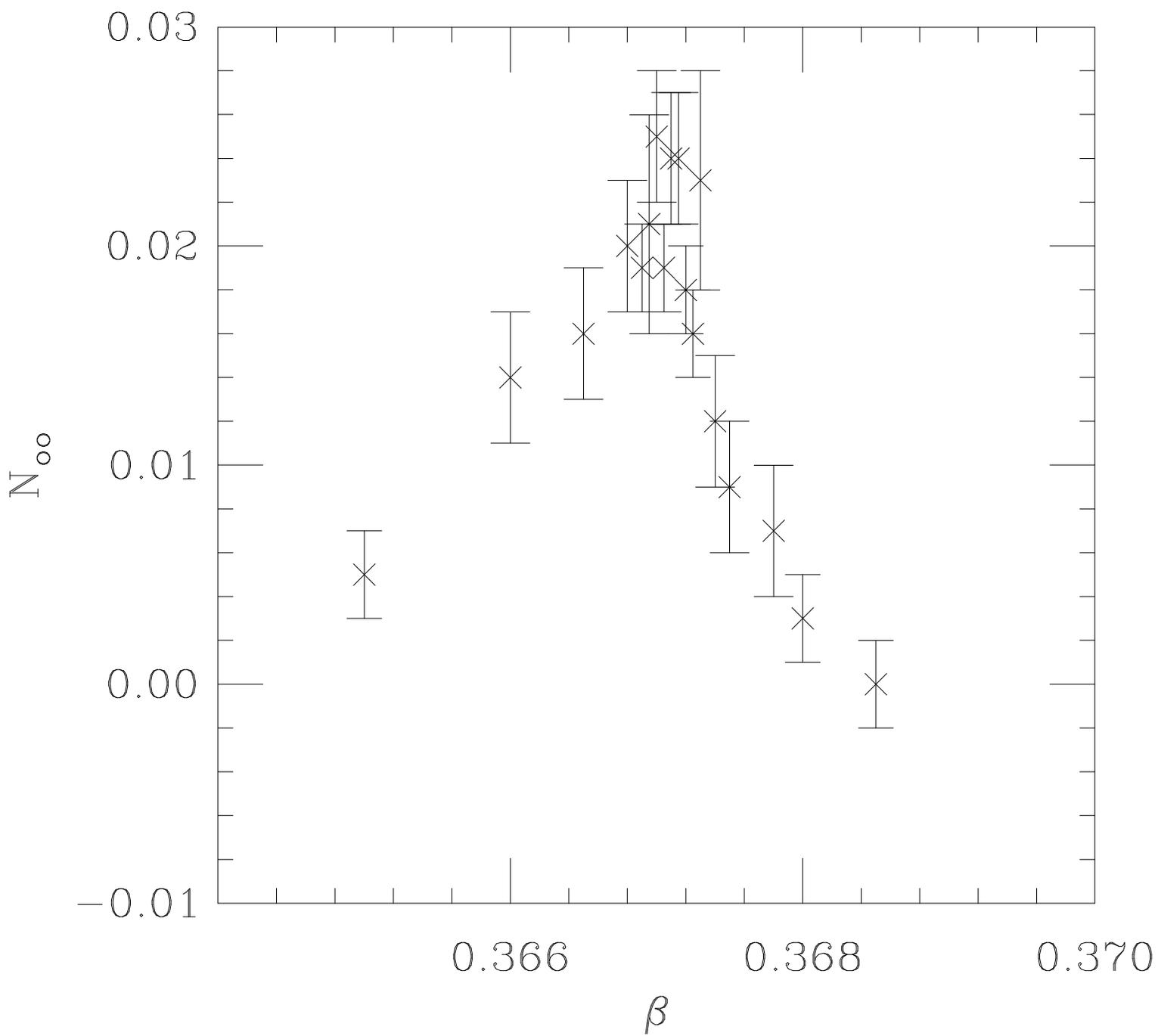
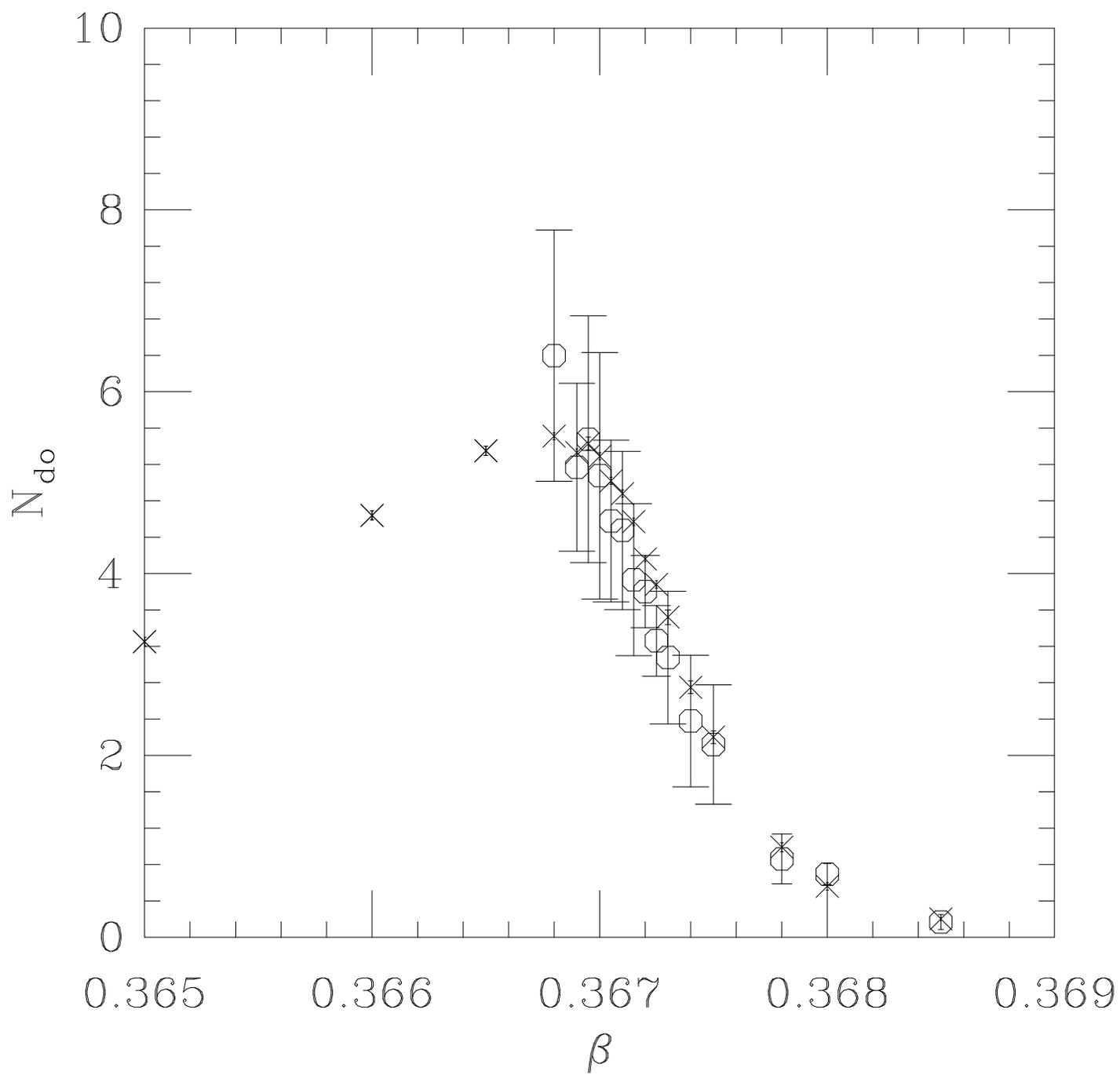
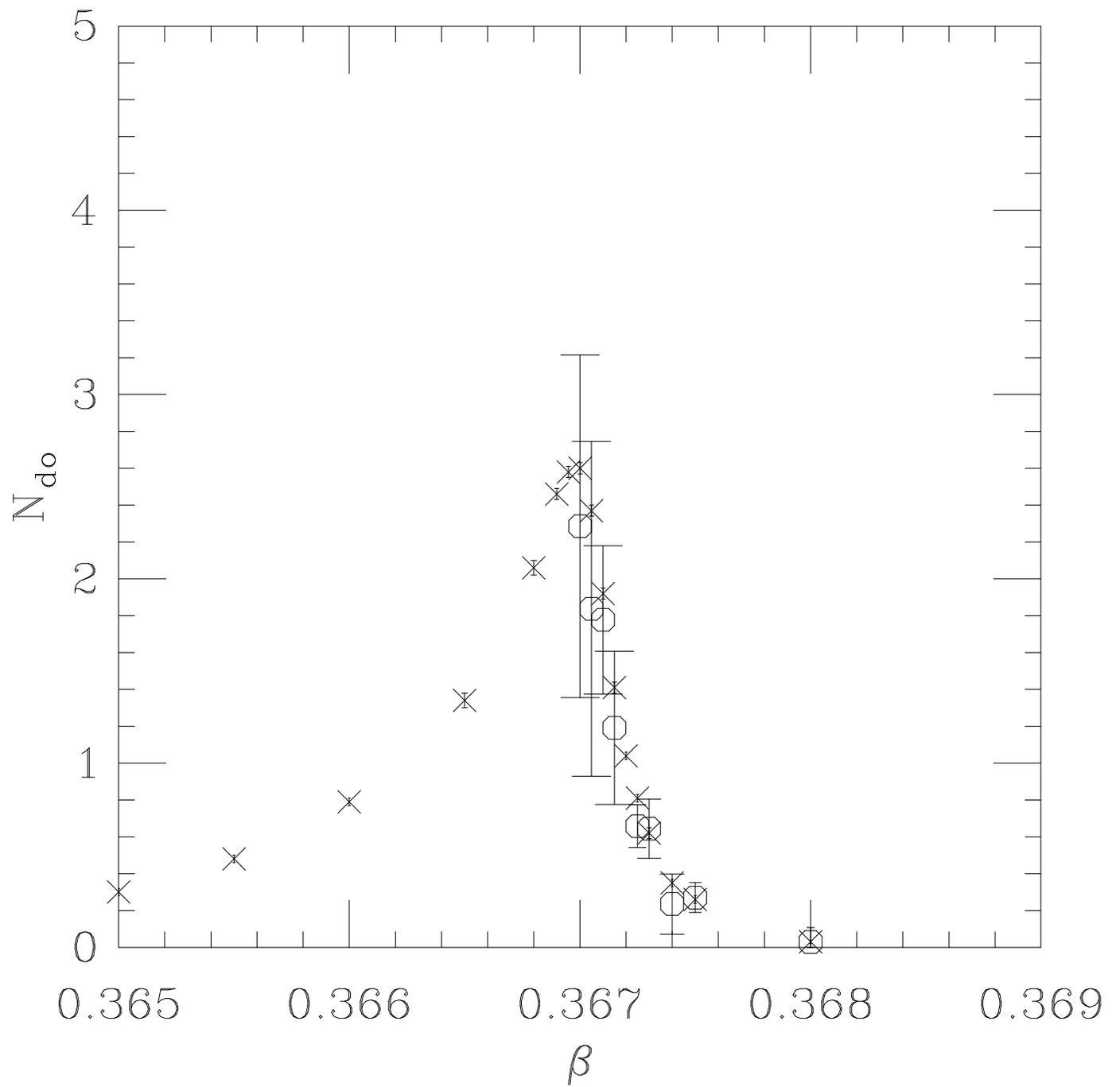
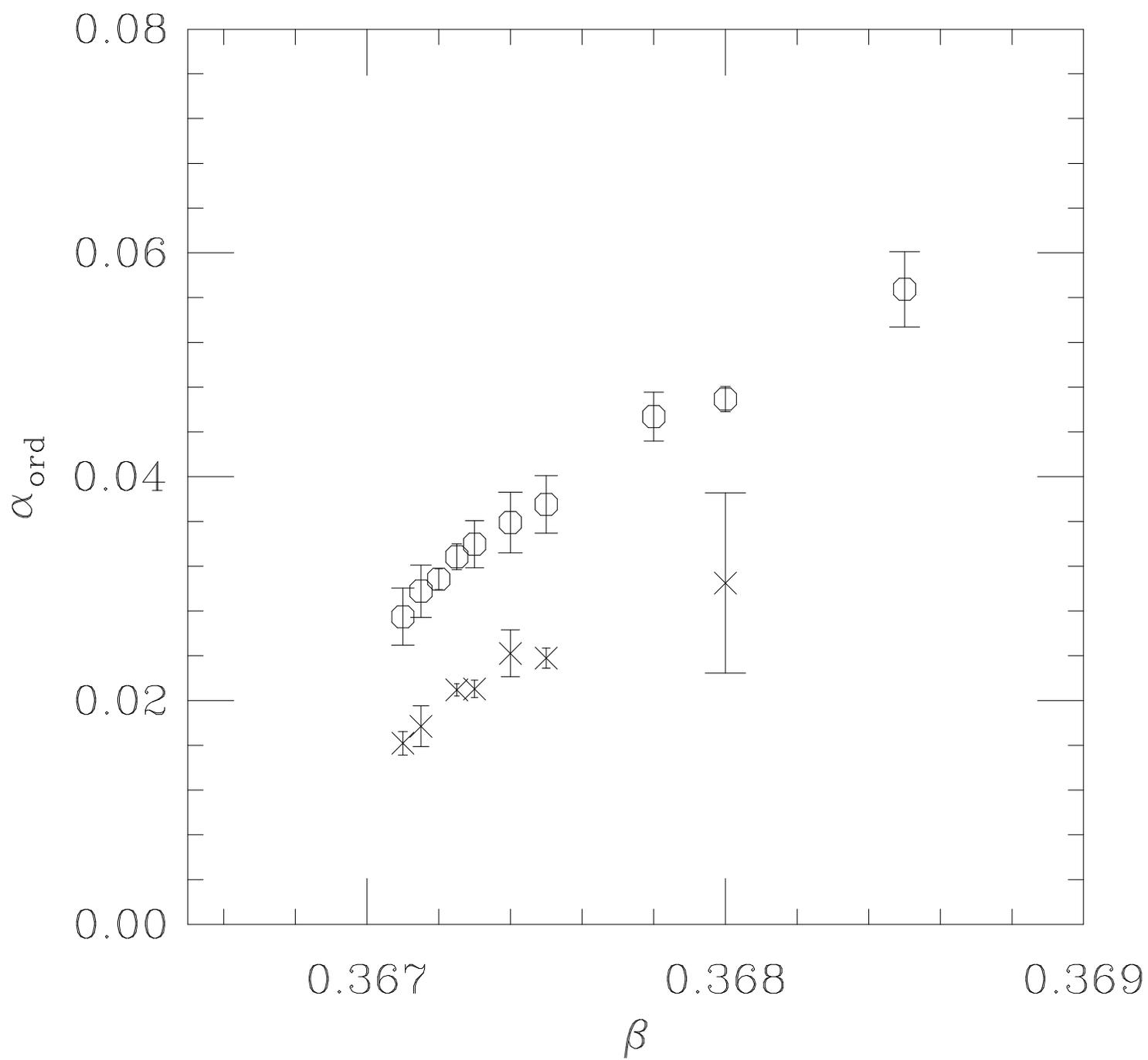
\end{document}